\documentclass[preprint,12pt]{elsarticle}

\usepackage{graphicx}
\usepackage{bm}
\usepackage{amsmath}
\usepackage{color}

\journal{Journal of Solid State Chemistry}

\begin{document} 

\begin{frontmatter}



\title{Pb$_5$Bi$_{24}$Se$_{41}$: A New Member of the Homologous Series
Forming Topological Insulator Heterostructures}


\author{Kouji Segawa, A. A. Taskin, Yoichi Ando$^\ast$}

\address{Institute of Scientific and Industrial Research, Osaka University \\
 8-1 Mihogaoka, Ibaraki, Osaka 567-0047, Japan}


\begin{abstract}
We have synthesized Pb$_5$Bi$_{24}$Se$_{41}$, which is a new member of
the (PbSe)$_5$(Bi$_2$Se$_3$)$_{3m}$ homologous series with $m$ = 4. This
series of compounds consist of alternating layers of the topological
insulator Bi$_2$Se$_3$ and the ordinary insulator PbSe. Such a
naturally-formed heterostructure has recently been elucidated to give
rise to peculiar quasi-two-dimensional topological states throughout the
bulk, and the discovery of Pb$_5$Bi$_{24}$Se$_{41}$ expands the
tunability of the topological states in this interesting homologous
series. The trend in the resistivity anisotropy in this homologous
series suggests an important role of hybridization of the topological
states in the out-of-plane transport.
\end{abstract}


\end{frontmatter}


\date{today}


%
%
%


\section{Introduction}

Topological insulators (TIs) are a new class of materials characterized
by a nontrivial topology of the Hilbert space spanned by the wave
functions of the occupied electronic states \cite{RMP_TI_10, Qi_RMP11,
Ando_Materials_Review}. They are expected to be useful for various
applications including high-frequency electronics, transparent
electrodes, spintronics, and quantum computations \cite{Moore,
PengNatChem, Cava}. Already a number of bulk materials have been found to
be three-dimensional (3D) TIs \cite{Ando_Materials_Review}, and they are
all narrow-gap semiconductors composed of heavy elements which cause
band inversions \cite{Ando_Materials_Review, Cava}. When the band
inversion occurs at an odd number of the high-symmetry points in the
Brillouin zone, the topological principle \cite{Ando_Materials_Review}
dictates that charge-conducting gapless states show up on the surface. 
Hence, 3D TIs are peculiar in that they consist of gapped bulk
states accompanied by gapless surface states of topological origin.
Syntheses of new 3D TI materials are of great current interest for
expanding our knowledge about the relationship between chemistry and
quantum-mechanical functionalities \cite{Cava}.

In this context, there is an intriguing class of materials whose crystal
structures realize a naturally-formed heterostructure consisting of
alternating layers of topological insulators and ordinary insulators
\cite{Ando_Materials_Review}; due to their constructions, such materials
can be considered to lie at the boundary between topologically trivial
and nontrivial materials. Specifically, in the homologous series of
Pb-Bi-Se ternary compounds expressed in the formula
(PbSe)$_5$(Bi$_2$Se$_3$)$_{3m}$ \cite{Kanatzidis_Pb_homologous_05,
Pb_homologous_2005, Shelimova_Pb_08}, which we call PSBS here, layers of
PbSe (which is an ordinary insulator) alternate with layers of
Bi$_2$Se$_3$ (which is a prototypical topological insulator
\cite{Xia_Nphys09}). In these series of compounds, it was inferred from
the band structure data obtained from angle-resolved photoemission
spectroscopy (ARPES) that each internal interface between PbSe and
Bi$_2$Se$_3$ layers harbor topological interface states, leading to the
situation that the whole bulk is filled with quasi-two-dimensional
states of topological origin \cite{Nakayama_PRL12}.

Interestingly, since the topological interface states appear on both
sides of each Bi$_2$Se$_3$ layer sandwiched by PbSe layers, the two
topological states at the top and bottom interfaces can hybridize {\it
within} each Bi$_2$Se$_3$ layer, leading to the opening of a gap in the
topological states \cite{Nakayama_PRL12}. The size of this hybridization
gap becomes smaller as the Bi$_2$Se$_3$ layer becomes thicker and the
hybridization becomes weaker. Based on the electronic structure data
obtained from ultrathin films of Bi$_2$Se$_3$ \cite{Zhang_NPhys10}, it is
expected that the hybridization is gone when the Bi$_2$Se$_3$ layer
becomes thicker than 5 nm. Note that the crystal structure of
Bi$_2$Se$_3$ consists of covalently-bonded Se-Bi-Se-Bi-Se quintuple
layers (QLs) that are interconnected by weak van der Waals force
\cite{Ando_Materials_Review}. The thickness of each QL is 0.9545 nm.
Thus, the hybridization of topological states in PSBS would disappear
when each Bi$_2$Se$_3$ layer contains more than 5 QLs. 

In the (PbSe)$_5$(Bi$_2$Se$_3$)$_{3m}$ homologous series, the parameter
$m$ gives the number of QLs included in each Bi$_2$Se$_3$ layer, and the
topological states becomes more robust for larger $m$. So far, syntheses
of compounds with $m$ = 1, 2, and 3 have been reported
\cite{Kanatzidis_Pb_homologous_05, Pb_homologous_2005, Shelimova_Pb_08,
Pb_Hetero_Ag_super}, and it is naively expected that, based on the phase
diagram of the PbSe-Bi$_2$Se$_3$ pseudo-binary system reported by
Shelimova {\it et al.} \cite{Shelimova_Pb_08}, compounds with larger $m$
would not be naturally synthesized because they are not
thermodynamically stable. Despite this naive expectation, we have
successfully synthesized $m$ = 4 compound based on a melt-growth
strategy in which the composition of the melt changes continuously as
the growth proceeds. The expansion of the available compounds in this
interesting homologous series is useful, because one can tune the
hybridization of the topological states by changing $m$, which is a
unique tunability among topological materials. In this paper, we report
the synthesis and analyses of the Pb$_5$Bi$_{24}$Se$_{41}$ [=
(PbSe)$_5$(Bi$_2$Se$_3$)$_{12}$, called PSBS $m$ = 4] compound, as well
as its anisotropic transport properties which support the
quasi-two-dimensional nature of its electronic states.

\section{Materials and methods}

\subsection{Experimental Methods}

The X-ray diffraction (XRD) analysis using 2$\theta$--$\theta$ scan is
performed with Rigaku Ultima-IV X-ray apparatus equipped with a
1D-detector. For single-crystal X-ray analysis, Rigaku Mercury CCD
system with graphite-monochromated MoK$\alpha$ radiation is used. The
in-plane transport properties of PSBS single crystals are measured by
standard ac six-probe method, which allows measurements of the in-plane
resistivity $\rho_{ab}$ and the Hall resistivity $\rho_{\rm H}$ at the
same time. By sweeping the magnetic field up to $\pm$2 T, the linear
slope of $\rho_{\rm H}$ vs. $B$ is obtained, from which the Hall
coefficient $R_{\rm H}$ is calculated. The out-of-plane resistivity
$\rho_{c^*}$ is measured by using a four-probe method, in which the
current and voltage contacts are made near the edges of the top and
bottom surfaces of a rectangular sample. In the present work,
$\rho_{ab}$ and $\rho_{c^*}$ are measured on the same samples to avoid
unnecessary complications, and the results from several samples are
compared to assess the reproducibility of the anisotropy measurements.

\subsection{Crystal Growth}

Single crystalline samples of PSBS with a series of $m$ values are grown
by a combination of modified Bridgman and self-flux methods. The growth
process is complicated, because these compounds do not exhibit congruent
melting \cite{Shelimova_Pb_08}. To grow high-quality single crystals,
the following steps are taken: High-purity raw material shots of Pb
(99.998\%), Bi (99.9999\%), and Se (99.999\%) are sealed in a quartz
tube that was heat-treated beforehand. For removing oxidization layers
formed in air on the raw shots of Pb and Bi, surface cleaning procedures
are performed; Pb shots are annealed in hydrogen atmosphere, and Bi
shots are washed with diluted HNO$_3$. Based on the phase diagram
reported by Shelimova {\it et al.} \cite{Shelimova_Pb_08}, we choose the
starting composition of PbSe:Bi$_2$Se$_3$ = 45:55 in molar ratio. The
raw materials are reacted and homogenized in a melt held at 1173 K for 6
h, and the crystal growth was fostered by slowly sweeping the
temperature from 1023 K to 923 K at a rate of 2 K/hour in a temperature
gradient of roughly 1 K/cm. At the beginning of the growth, $m$ = 1
crystals are formed at the low-temperature end of the melt, which causes
the composition of the melt to become PbSe-poor. As the temperature is
lowered and the crystals growth proceeds, the melt becomes more and more
PbSe-poor and the composition of the grown crystals changes from $m$ = 1
to higher $m$ values.

\section{Results : Grown Crystals}

After cutting the grown boule [{Fig. 1}a] with a wire saw,
cleavable single crystals can be separated. The phase of each piece is
determined by the 2$\theta$--$\theta$ XRD analysis
on a cleaved surface. Even when the appearance of the surface is clean
and shiny, the XRD profile can sometimes contain peaks from multiple
phases; we discard such multi-phase crystals and only use those crystals
that show single-phase peaks on both top and bottom surfaces for further
characterizations. The largest size of the single-phase crystal we
obtained was roughly $2 \times 1 \times$0.2 mm$^3$. The XRD pattern
confirmed that the cleavage plane is along the $ab$-plane. Note that,
because the crystal structures of the PSBS homologous series are
monoclinic \cite{Pb_Hetero_Ag_super}, the $c$-axis is {\it not}
perpendicular to the $ab$-plane, and hence the direction normal to the
$ab$ plane is called $c^\ast$-axis. This means that the
2$\theta$--$\theta$ XRD profile on a cleaved surface of a PSBS single
crystal reflects the periodicity along the $c^\ast$-axis, which
corresponds to $c \sin\beta$ with $c$ and $\beta$ the $c$-axis lattice
constant and the angle between the $c$-axis and the $ab$-plane,
respectively. The periodicities (i.e. the $c\sin\beta$ values) for $m$ =
1, 2, and 3 are obtained in the present experiments to be 1.5795(19),
2.5287(11), and 3.4753(15) nm, respectively [Fig. 1b-d and {Table
1}] \cite{Notice_for_c}; these results are consistent with those
previously reported \cite{Shelimova_Pb_08, Pb_Hetero_Ag_super}. However,
in addition to these three phases, we found a new phase of unknown XRD
profile, as shown in Fig. 1e. The crystals of this new phase presents
a shiny cleavage plane, and the appearance is indistinguishable from
other PSBS phases nor the Bi$_2$Se$_3$ compound. Nevertheless, from the
XRD profile, the periodicity perpendicular to the cleavage plane is
calculated to be 4.4274(15) nm, which is very long and is obviously
different from any known phase in this Pb-Bi-Se ternary system. The
detailed XRD data for all four phases are presented in the Supplementary
Materials.

\begin{figure}
\begin{center}
\includegraphics[width=11.5cm,clip]{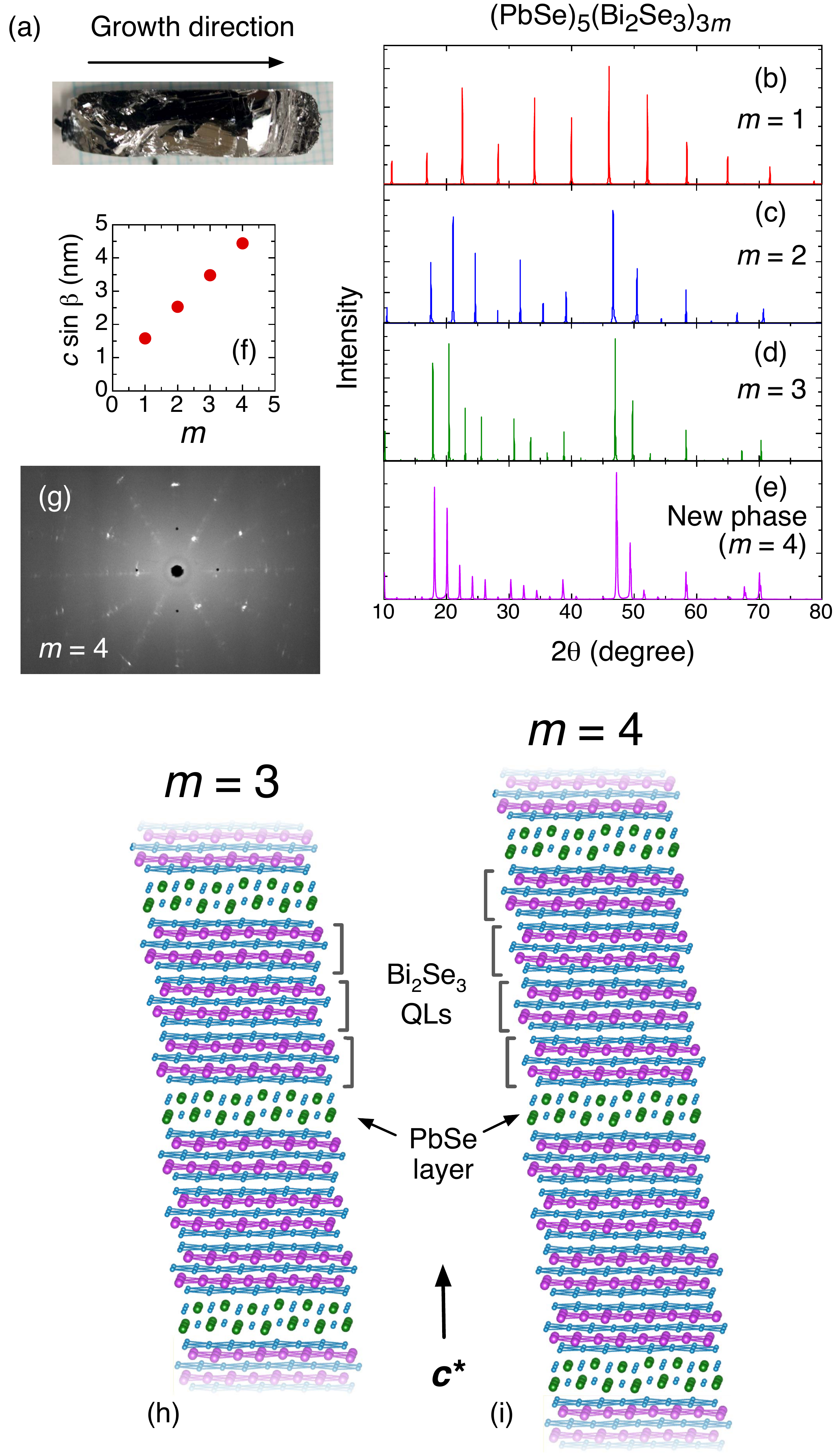}
\caption{
(a) Photograph of the boule grown from the raw composition of 
PbSe:Bi$_2$Se$_3$ = 45:55.
\
(b-e) 2$\theta$--$\theta$ XRD profiles taken on 
cleaved surfaces of single-crystal samples of the PSBS homologous series
with various $m$ values: (b)  
$m$ = 1; (c) $m$ = 2; (d) $m$ = 3; (e) new phase, $m$ = 4.
(f) Plot of $c\sin\beta$ vs. $m$, where the data point from the new phase 
is assigned to $m$ = 4.
(g) A Laue picture taken on the cleaved surface of a single crystal of the 
$m$ = 4 phase.
(h, i)
Schematic pictures of the crystal structures drawn \cite{VESTA}
for the $m$ = 3 and 4 phases of the PSBS homologous series. Because of the 
monoclinic structure, the direction normal to the cleavage plane
($ab$-plane) is denoted as $c^*$-axis.}
\label{Crystal_plot}
\end{center}
\end{figure}

\begin{table}[b]
\centering 
\begin{tabular}{c c c} 
\hline\hline 
Composition & \,\,\, $d_{\rm PbSe-PbSe}$ \,\,\, &  $\Delta d_{\rm PbSe-PbSe}$  \\ 
(PbSe)$_5$(Bi$_2$Se$_3$)$_{3m}$  & ($=c\sin\beta$; nm) & (nm) \\ 
[0.5ex] 
\hline 
$m$ = 1 & 1.5795(19) & -- \\
$m$ = 2  & 2.5287(11) & 0.9492  \\
$m$ = 3  & 3.4753(15) & 0.9466  \\
New phase ($m$ = 4) & 4.4274(15) & 0.9521   \\
[1ex] 
\hline 
\end{tabular}
\caption{$\Delta d_{\rm PbSe-PbSe}$ is the increment in $d_{\rm PbSe-PbSe}$ 
(= $c\sin\beta$)
when $m$ increases by one.} 
\end{table}

In the PSBS compound, an increase in $m$ corresponds to an increase in
the thickness of the Bi$_2$Se$_3$ layer. In fact, when $m$ increases by
one from $m$ = 1 to 2 and from $m$ = 2 to 3, the $c\sin\beta$ value
increases by 0.9492 and 0.9466 nm, respectively; these values are close
to 0.9545 nm, the thickness of the QL unit of bulk Bi$_2$Se$_3$. In this
respect, the difference in $c\sin\beta$ between the $m$ = 3 phase and
the new phase is 0.9521 nm, which is also close to the QL-unit thickness
of Bi$_2$Se$_3$, suggesting that the new phase corresponds to the $m$ =
4 phase of the homologous series. This systematics is graphically
confirmed by plotting $c\sin\beta$ vs. $m$ (assuming that the new phase
is $m$ = 4), which clearly presents a linear relation [Fig. 1f].
Figure 1g shows a Laue picture taken on the $ab$-plane of a single
crystal of the
$m$ = 4 phase. The same pattern is obtained from any spot on the surface,
which confirms the single-crystal nature of the sample.
The Laue picture clearly shows three-fold symmetry and is very similar to that
of the $m$ = 2 and 3 phases. 
We show schematic pictures of the crystal structures for $m$ = 3 and 4 in
Fig. 1h and 1i;
these pictures are drawn by expanding the crystal
structure data for $m$ = 2 \cite{Pb_Hetero_Ag_super}.

We have further analyzed the crystal structure of the new phase with
single-crystal X-ray apparatus with CCD imaging. The space group is
identified to be monoclinic P2 with the Laue class 2/$m$. The lattice
parameters are $a = 2.167(3)$ nm, $b=0.4123(5)$ nm, $c=4.431(8)$ nm,
$\beta=93.35(5)^\circ$, and $c\sin\beta=4.424(9)$ nm. The values of $a$
and $b$ are close to those reported for $m$ = 1 and 2
\cite{Pb_Hetero_Ag_super}, and $c\sin\beta$ is in good agreement with
that obtained from the $2\theta$--$\theta$ analysis. It is useful to
note that $\beta$ is closer to 90$^\circ$ than that for smaller $m$
\cite{Shelimova_Pb_08}, and this is naturally explained by the elongated
$c$ unit while keeping $a$ and $b$ units to be essentially unchanged.
Further refinement was tried both for powder and single-crystal XRD data,
but it was difficult 
because the crystals are relatively soft and easily
deformed, and also because the PSBS system appears to be metastable and
the $m$ = 4 homologous phase disappears upon grinding.
More detailed information on this X-ray analysis is presented in the
Supplementary Materials.

\begin{table}[b]
\centering 
\begin{tabular}{c c c c c c} 
\hline\hline 
$m$ & Nominal & Pb &  Bi &  Pb+Bi  & Se \\
& \,\,\, composition \,\, & & & & \\
[0.5ex] 
\hline 
1 & Pb$_5$Bi$_6$Se$_{14}$ & 3.38(10) & 7.47(3) & 10.9(7) & 14  \\
2 & Pb$_5$Bi$_{12}$Se$_{23}$  & 3.25(6) & 13.6(2) & 16.8(1) & 23 \\
3 & Pb$_5$Bi$_{18}$Se$_{32}$  & 2.77(11) & 20.2(4) & 23.0(3) & 32  \\
4 & Pb$_5$Bi$_{24}$Se$_{41}$  & 3.26(7) & 25.5(2) & 28.8(1) & 41   \\
 [1ex] 
\hline 
\end{tabular}
\caption{The data are obtained from ICP-AES analyses.
The selenium composition is fixed to be 5+9$m$.} 
\end{table}

To quantitatively determine the actual compositions of the grown
crystals, inductively coupled plasma atomic emission spectroscopy
(ICP-AES) analysis is employed. The results for all the grown phases,
including the new phase, are shown in {Table 2}, in which the Se
contents are fixed to the stoichiometric values calculated from the
formula (PbSe)$_5$(Bi$_2$Se$_3$)$_{3m}$; namely, the Se contents are set
to be 14, 23, 32, and 41 for $m$ = 1, 2, 3, and 4, respectively, and the
Pb and Bi contents are calculated against these values, because the
ICP-AES analysis is best at giving relative ratios between constituent
elements. One can see that the Pb content is much smaller than the
nominal value of 5 in all the compounds. In contrast, the Bi content is
always larger than the nominal values. These results point to the
existence of high densities of Bi antisite defects, Bi$^{\bullet}_{\rm
Pb}$, and in fact, the summed values of Pb and Bi contents (shown as
``Pb+Bi'' in Table 2) are in reasonable agreement with the nominal
values. Therefore, the cation/anion ratios obtained by the ICP-AES
analyses support the conclusion that all four phases are members of the
PSBS homologous series.

\begin{table}[b]
\centering 
\begin{tabular}{c c c c} 
\hline\hline 
$m$ & \,\, $x$ (Pb) \,\, &  \,\, $y$ (Bi) \,\, &  \,\, $\Delta q$ (per mol)\\ 
[0.5ex] 
\hline 
1 & 0.68(2) & 0.29(1) & 1.0--1.3  \\
2 & 0.65(1) & 0.31(3) & 0.8--1.6 \\
3 & 0.55(2) & 0.45(8) & 1.1--3.3  \\ 
4 & 0.65(1) & 0.30(3) & 0.7--1.4  \\
 [1ex] 
\hline 
\end{tabular}
\caption{The $x$ and $y$ values quantify the off-stoichiometry in terms of
the formula
(Pb$_x$Bi$_y$Se)$_5$(Bi$_2$Se$_3$)$_{3m}$ determined from the ICP-AES
analysis for each $m$. $\Delta q$ gives the charge imbalance per mole
due to the off-stoichiometry calculated by $\Delta q = 5(2x+3y-2)$; in
this table, because of the errors in $x$ and $y$ values, the ranges of
$\Delta q$ dictated by the errors are shown.}
\end{table}

The above ICP-AES results indicate that the actual chemical formula for
PSBS had better be expressed as
(Pb$_{x}$Bi$_y$Se)$_5$(Bi$_2$Se$_3$)$_{3m}$
(note that the solubility of Pb into Bi$_2$Se$_3$ is reported to be
negligible \cite{Aliev}).
We show in {Table 3}
the $x$ and $y$ values in this formula for $m$ = 1 -- 4 calculated from
the ICP-AES results. The Bi$^{\bullet}_{\rm Pb}$ antisites are donors to
provide $n$-type carriers, and thus one can calculate the expected
$n$-type carrier density from the charge imbalance $\Delta q =
5(2x+3y-2)$ per mol; this formula is based on the valence states
Pb$^{2+}$, Bi$^{3+}$, and Se$^{2-}$. Positive $\Delta q$ corresponds to
$n$-type. The expected carrier density $n_e$ is obtained by dividing
$\Delta q$ by the unit-cell volume, and for $m$ = 4, one obtains $n_e =
2 \times 10^{20} - 4 \times 10^{20}\, {\rm cm}^{-3}$.

\section{Results : Transport Properties}

{Figure 2}a shows the temperature dependences of $\rho_{ab}$
for five samples of PSBS $m$ = 4 single crystals
obtained from the same boule. The temperature dependence is metallic,
and the absolute value of $\rho_{ab}$ varies by about 40\% between
samples. The temperature dependences of $R_H$ for
the same set of samples are shown in Fig. 2b; the variation of $R_H$
between samples is again about 40\% at room temperature. It is useful to
note that the data in Fig. 2b tell us that the sample-to-sample
variation in $\rho_{ab}$ is caused essentially by a change in the
carrier density, which is presumably related to different levels of Bi
antisite defects in different pieces of crystals. In fact, as shown in
Fig. 2c, the Hall mobility $\mu_H$ calculated from $\rho_{ab}$ and
$R_H$ for those samples converges at room temperature, indicating that
the level of electron scattering is unchanged among samples at room
temperature. The carrier density $n_H$ calculated from $R_H$ varies from
$1\times 10^{20} \,{\rm cm}^{-3}$ to $3\times 10^{20}\,{\rm cm}^{-3}$,
which is smaller than that expected from the density of Bi antisites by
roughly a factor of two. A major part of this discrepancy is probably
due to the Hall factor, which accounts for the difference between the
actual carrier density and that calculated from $1/(eR_H)$, because the
latter can vary due to the multiband effects and the relaxation-time
anisotropy.

\begin{figure}[t]
\begin{center}
\includegraphics[width=12.5cm,clip]{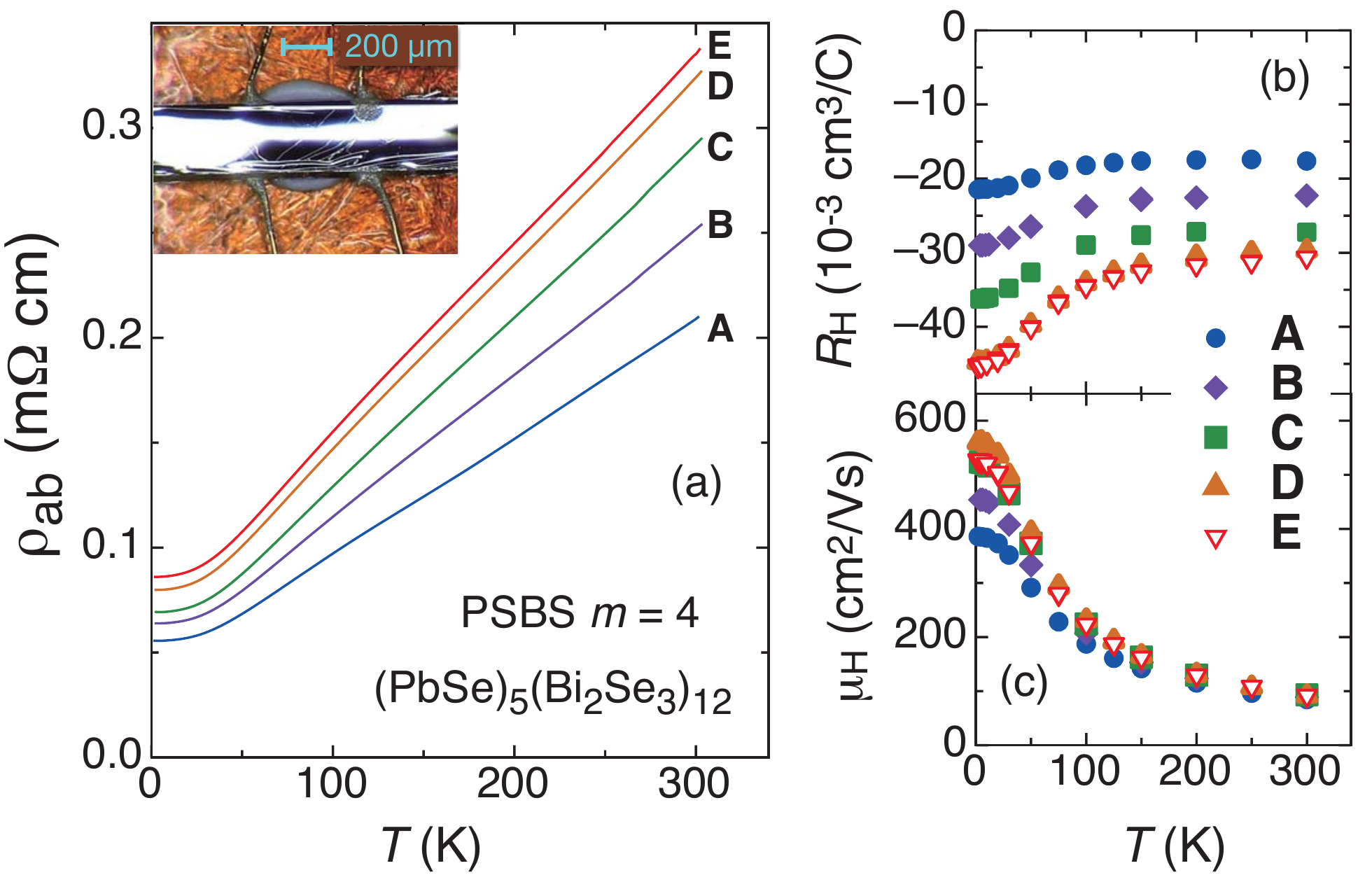}
\caption{
Transport properties of the new compound, Pb$_5$Bi$_{24}$Se$_{41}$ (PSBS
$m$ = 4), measured on five different pieces of single crystals (A--E).
Panels (a)-(c) present the temperature dependences of (a) $\rho_{ab}$,
(b) $R_H$, and (c) $\mu_H$. The inset of panel (a) shows a photograph of 
sample C with four voltage contacts.}
\label{Transport_plot}
\end{center}
\end{figure}

The out-of-plane resistivity $\rho_{c^*}$ is useful for evaluating the
anisotropy and the dimensionality of the charge transport. In {Fig. 3}a,
the temperature dependences of $\rho_{c^*}$ for the same set
of samples as those in Fig. 2 are shown. The magnitude of $\rho_{c^*}$
is of the order of 10 m$\Omega$cm, which is large for a degenerate
semiconductor with $n_e \simeq 1\times 10^{20} \,{\rm cm}^{-3}$. The
temperature dependence is either weakly metallic or almost temperature
independent, from which it is difficult to judge whether the transport
along the $c^*$-axis is coherent. Nevertheless, as shown in Fig. 3b,
the anisotropy ratio $\rho_{c^*}/\rho_{ab}$ grows with lowering
temperature and reaches 75--170 at low temperature, which points to the
quasi-two-dimensional nature of the transport \cite{cuprate_anisotropy}.
For comparison, we show in Fig. 3c-e the data of $\rho_{ab}$,
$\rho_{c^*}$, and $\rho_{c^*}/\rho_{ab}$ for two samples of PSBS $m$ = 2
single crystals. One can see that $\rho_{ab}$ is comparable to that of
the $m$ = 4 phase, but $\rho_{c^*}$ (and hence $\rho_{c^*}/\rho_{ab}$)
is much smaller in the $m$ = 2 phase. Thus, the present results suggest
that the transport anisotropy increases with increasing $m$ in the PSBS
homologous series. (We note that it has so far been difficult to obtain
sufficiently large single crystals of the $m$ = 3 phase, and hence we do
not have reliable data for the resistivity anisotropy for $m$ = 3.)

\begin{figure}[t]
\begin{center}
\includegraphics[width=12.5cm,clip]{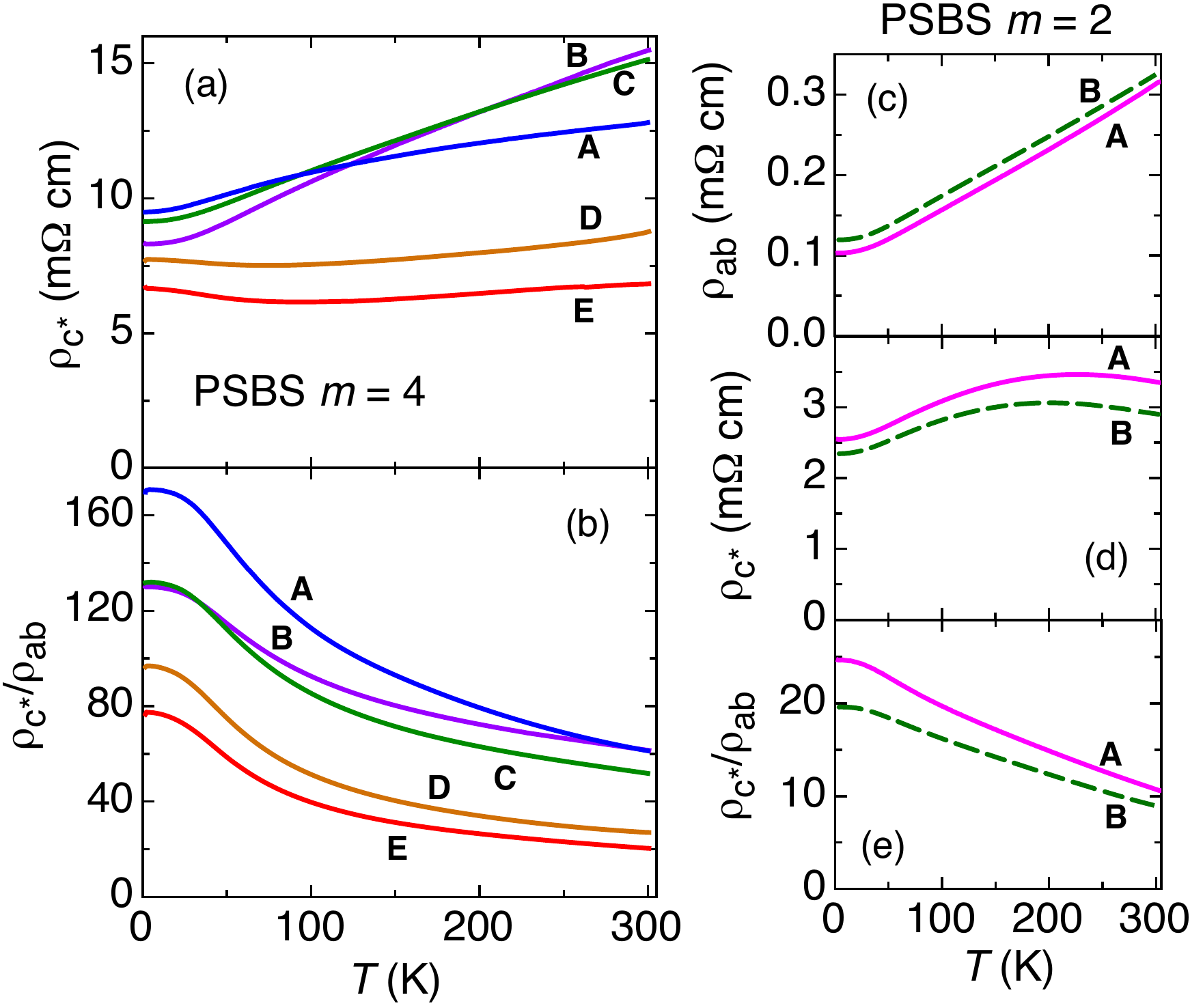}
\caption{
(a-b) Temperature dependences of (a) $\rho_{c^*}$ and (b) $\rho_{c^*}/\rho_{ab}$ 
measured on the same $m$ = 4 samples (A--E) as those shown in Fig. 2.
(c-e) Temperature dependences of (c) $\rho_{ab}$, (d) $\rho_{c^*}$, 
and (e) $\rho_{c^*}/\rho_{ab}$ measured on two $m$ = 2 samples (A and B).}
\label{Anisotropy_plot}
\end{center}
\end{figure}

\section{Discussions}

To understand the implications of the difference in the resistivity
anisotropy between $m$ = 2 and 4, it is useful to remember that in bulk
Bi$_2$Se$_3$,
which can be viewed as the $m \rightarrow \infty$ limit of the PSBS
homologous series,
the anisotropy is as small as $\sim$3
\cite{Bi2Se3_anisotropy}, and this is because Bi$_2$Se$_3$ has a 3D
ellipsoidal Fermi surface. Therefore, the {\it increase} in the
anisotropy for larger $m$ cannot be an indication that the system is
approaching the $m \rightarrow \infty$ limit. In other words, the
present result on the resistivity anisotropy strongly supports the
conjecture that PbSe layer in this homologous series works as a block
layer for the charge transport. Nevertheless, the volume density of the
PbSe layer becomes {\it smaller} for larger $m$, and hence the larger
$\rho_{c^*}$ value at $m$ = 4 cannot be due simply to the presence of
PbSe layers as series resistors.

One possibility to understand the trend in the resistivity anisotropy is
to consider that the level of hybridization between the top and bottom
interface states of topological origin within each Bi$_2$Se$_3$ layer
plays an important role in determining the out-of-plane transport. In
such a case, a thicker Bi$_2$Se$_3$ layer (i.e. a larger $m$) means
weaker hybridization, which would lead to smaller transfer integral
between top and bottom interface states within the Bi$_2$Se$_3$ layer,
resulting in a larger $c^*$-axis resistivity. Although speculative, this
possibility suggests the importance of the topological interface state
in the out-of-plane transport in the PSBS homologous series.

We now discuss the chemical reason why it was possible to synthesize the
$m$ = 4 phase in the present work. According to the phase diagram of the
PbSe-Bi$_2$Se$_3$ pseudo-binary system reported by Shelimova {\it et
al.} \cite{Shelimova_Pb_08}, there are only three stable phases,
Pb$_5$Bi$_6$Se$_{14}$ ($m$ = 1), Pb$_5$Bi$_{12}$Se$_{23}$ ($m$ = 2), and
Pb$_5$Bi$_{18}$Se$_{32}$ ($m$ = 3), which respectively form below 993,
973, and 948 K. The phase diagram also suggests that if the
Bi$_2$Se$_3$:PbSe ratio is higher than that corresponds to $m$ = 3, a
mixed phase solidifies below 930 K. However, this phase diagram was
constructed from only a finite number of compositional data points, and
it is possible that there is a narrow compositional window in which the
$m$ = 4 phase preferentially grows below some temperature that lies
between 930 and 948 K. In our growth method for this homologous series,
the composition of the melt changes continuously toward a higher
Bi$_2$Se$_3$:PbSe ratio as the temperature is lowered and the crystal
growth proceeds, which must have brought the system to hit the right
composition/temperature window for the $m$ = 4 growth. Hence, our method
is suitable for the growth of rare compounds in a homologous series,
although it is difficult to obtain large single crystals for each phase.

Finally, it would be useful to elaborate on the difference in the
crystal structures of the homologous series obtained for Pb-Bi-Se and
Pb-Bi-Te ternary systems. The Pb-Bi-Te ternary system also gives rise to
a homologous series, but its construction is fundamentally different
from that obtained for the Pb-Bi-Se ternary system. Namely, in the case
of tellurides, although the composition formula is superficially similar
to the selenides and is written as (PbTe)$_m$(Bi$_2$Te$_3$)$_n$ ($m$ and
$n$ are integers), in this homologous series ``PbTe'' is incorporated
into the original rhombohedral structure of Bi$_2$Te$_3$, forming a
covalently-bonded Te-Bi-Te-Pb-Te-Bi-Te septuple-layer unit
\cite{SoumaPb124_PRL12, Pb147_NC12}. Hence, in the telluride homologous
series, the rhombohedral symmetry of the crystal structure which stems
from a cubic structure distorted along (111) direction is essentially
unchanged from that of Bi$_2$Te$_3$. Importantly, the $(m,n) = (1,1)$
phase of this telluride homologous series, PbBi$_2$Te$_4$, which simply
consists of a stack of the Te-Bi-Te-Pb-Te-Bi-Te septuple-layers, is
known to be a topological insulator \cite{SoumaPb124_PRL12, Pb147_NC12};
hence, the telluride homologous series is a heterostructure of two
topological insulators (i.e. PbBi$_2$Te$_4$ and Bi$_2$Te$_3$) and the
topological states show up only on the surrounding surface. In contrast,
in the selenide homologous series \cite{Kanatzidis_Pb_homologous_05,
Pb_homologous_2005, Shelimova_Pb_08, Pb_Hetero_Ag_super}, PbSe is not
incorporated into the rhombohedral structure of Bi$_2$Se$_3$ and the
rock-salt structure of bulk PbSe partly remains, forming the PbSe layer
of square symmetry; this layer alternates with the Bi$_2$Se$_3$ layer of
hexagonal symmetry, and the resulting structure is no longer
rhombohedral. As is already emphasized, this selenide homologous series
is a heterostructure of a topological insulator and an ordinary
insulator, and hence its electronic nature is fundamentally different
from that of the telluride homologous series.

\section{Conclusions}

We have synthesized a new compound in the
(PbSe)$_5$(Bi$_2$Se$_3$)$_{3m}$ homologous series with $m$ = 4, which
became possible by using a melt-growth method in which the composition
of the melt changes continuously to allow the whole series of compounds
to grow in a sequential manner. This new compound,
Pb$_5$Bi$_{24}$Se$_{41}$, is a member of the unique family of
topological materials in which alternating layers of the topological
insulator Bi$_2$Se$_3$ and the ordinary insulator PbSe form natural
heterostructures, leading to the appearance of topological interface
states throughout the bulk. The trend in the $c^*$-axis resistivity and
the resistivity anisotropy with increasing $m$ values suggests that the
hybridization of the topological interface states plays an important
role in the out-of-plane transport in this compound.

\vspace{4mm}
\noindent {\bf Acknowledgment}\\ \\
We acknowledge the Comprehensive Analysis Center, ISIR, Osaka
University, for technical supports in the CCD X-ray diffraction
analysis. We also thank T. Toba for technical assistances. This work was
supported by JSPS (KAKENHI 25220708, 24540320, and 25400328), MEXT
(Innovative Area ``Topological Quantum Phenomena" KAKENHI 22103004), and
AFOSR (AOARD 124038).



\begin{thebibliography}{10}
\bibitem{Notice_for_c} Strictly speaking, the actual $c$ value must be doubled for $m$ = 2, 
because the unit cell is twice as large in this phase along $c$
\cite{Pb_Hetero_Ag_super}.
\end{thebibliography}
\end{document}